\documentclass[]{spie}  

\usepackage{amsmath,amsfonts,amssymb}
\usepackage{graphicx}
\usepackage{subfig}
\usepackage[colorlinks=true, allcolors=blue]{hyperref}

\title{Development of the Fabry-Perot interferometers for the HIRMES spectrometer on SOFIA}

\author[a]{Greg Douthit}
\author[a]{Gordon Stacey}
\author[a]{Thomas Nikola}
\author[a]{Chuck Hendersen}
\author[a]{George Gull}
\author[a]{Kayla Rossi}
\author[b]{Alexander Kutyrev}
\author[b]{Harvey Moseley}

\affil[a]{Cornell University, Ithaca, NY 14853, USA}
\affil[b]{NASA Goddard Space Flight Center, Greenbelt, MD 20771, USA}

\authorinfo{Further author information: (Send correspondence to Greg Douthit)\\: E-mail: jgd86@cornell.edu, Telephone: 770-778-4436\\}

\begin{document} 
\maketitle

\begin{abstract}
HIRMES is a 
far-infrared spectrometer that was chosen as the third generation instrument for NASA's SOFIA airborne observatory. HIRMES promises background limited performance in four modes that cover the wavelength range between 25 and 122 $\mu$m. The high-spectral resolution ($R \approx 10^5$) mode is matched to achieve maximum sensitivity on velocity-resolved lines to study the evolution of protoplanetary disks. The mid-resolution ($R \approx 12, 000$) mode is suitable for high sensitivity imaging of galactic star formation regions in, for example, the several far-infrared fine structure lines. The low-resolution ($R \approx 2000$) imaging mode is optimized for spectroscopic mapping of far-infrared fine structure lines from nearby galaxies, while the low resolution ($R \approx 600$) grating spectrometer mode is optimized for detecting dust and ice features in protostellar and protoplanetary disks. Several Transition Edge Sensed (TES) bolometer arrays will provide background limited sensitivity in each of these modes. To optimize performance in the various instrument modes, HIRMES employs eight unique fully-tunable cryogenic Fabry-Perot Interferometers (FPIs) and a grating spectrometer. Here we present the design requirements and the mechanical and optical characteristics and performance of these tunable FPI as well as the control electronics that sets the mirror separation and allows scanning of the FPIs.
\end{abstract}


\section{INTRODUCTION AND SCIENCE JUSTIFICATION}
\label{sec:intro}  

The High Resolution Mid Infrared Spectrometer (HIRMES) is the third generation science instrument for NASA's SOFIA. The development of HIRMES is led by Goddard Space Flight Center with Cornell University developing and providing the series of cryogenic scanning Fabry-Perot Interferometers described here.  HIRMES's primary science goal
is to investigate the physics and chemistry of protoplanetary disks and planet formation, using velocity resolved spectroscopy in the mid- to far-infrared spectral lines of HD, H\textsubscript{2}O, and [OI] and low resolution spectroscopy of solid state water ice features. These lines trace important building blocks in the formation of planets and the development of life.
\subsection{Protoplanetary Disks}
\begin{figure} [ht]
   \begin{center}
   \includegraphics[height=5cm]{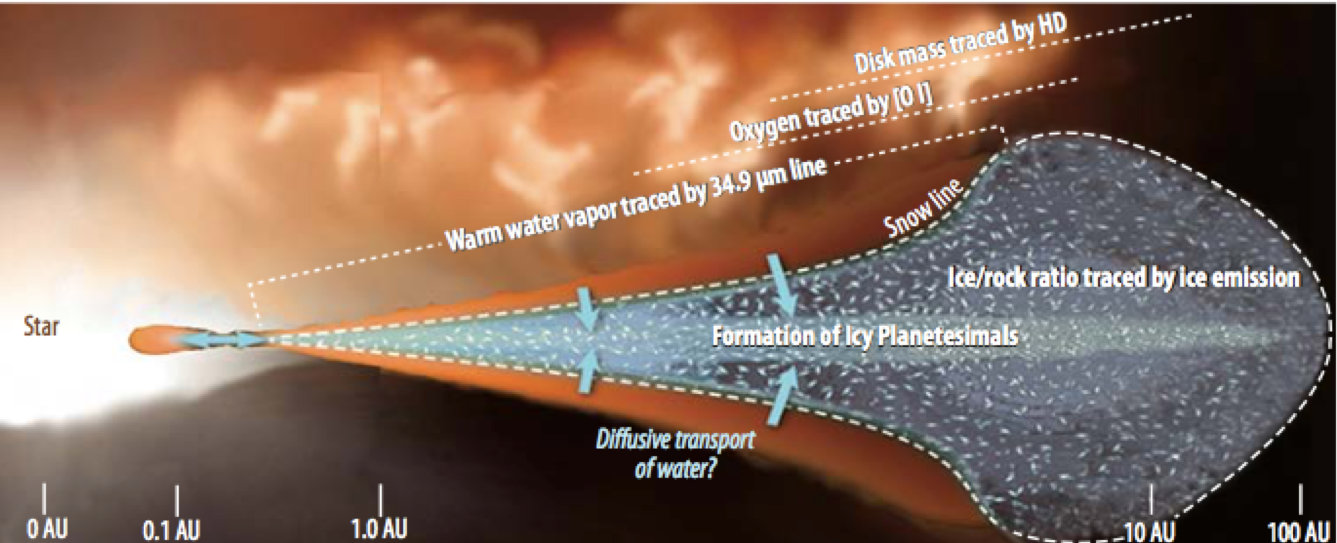}
   \end{center}
    \caption[disk] 
   {\label{waterdist} Model of water distribution in a protoplanetary disk. HIRMES will help determine the distribution of water looking for depletion in the hot layers above the snowline and investigate the settling of solid particles to the mid-plane.}	
 \end{figure}

Stars form from condensations in dense molecular cloud cores. During the final phases accretion disks form, enveloping and feeding the pre-main sequence stars. Some fraction of these disks also provide the raw material for the formation of planets.  Fig. \ref{waterdist} illustrates the morphology of this early phase of planet formation. It is thought that most planets form within a few tens of AU of the star in a disk of cold gas, dust and ice. The largest disk ice/rock particles and most dense gas and dust will naturally accumulate near the disk mid-plane, enveloped by a gas/dust ice/rock envelope extending perhaps an AU in scale height above the radially flaring plane. Material in the disk is shielded from the radiation of the protostar by dust. Where the shielding is high, ices will form, but towards the surface of the disk, temperatures will rise so that ices become gases. The region where water ice becomes water vapor is call the snow line. It is determined by the thermal balance of lessening of the stellar radiation field through shielding by dust and geometric fall-off.
Within the denser regions of the disk near its mid-plane, dust and ice grains will collide and occasionally stick together slowly growing larger as more matter falls onto the inner plane from the envelope. At size-scales greater than about 1 km in diameter, these solid bodies are called planetesimals - the seeds of future planets. Planetesimals will continue to grow through accretion and collisional coalescence until they reach size-scales of order 1000 km diameter where runaway gravitational accretion can take over forming \textit{proto-planets}. During the planet formation process, the young star will often go through stages of outflow and/or high stellar winds which will eventually disperse the protoplanetary disk limiting the mass of the planetary system. A key question is what fraction of the original protoplanetary gas and dust mass is incorporated into the planetary system, and what drives the elemental abundance ratios found in various types of planets such as terrestrial (rocky) planets and gas giants. 

Protostellar and protoplanetary disks are probed with a variety of tracers using state-of-the-art facilities across the electromagnetic spectrum including, for example optical/near-IR work on the Hubble Space telescope, mid- and far-IR work using the Spitzer and Herschel space telescopes, and submillimeter and millimeter wave interferometry with the ALMA interferometer. However these observations leave an incomplete story. Optical observations cannot probe deeply into the protoplanetary disks since optical radiation from the inner disk will be obscured by dust both in the outer disk and the enveloping molecular cloud. Mid and far-IR radiation can escape from the inner regions, but the spatial resolution of Spitzer and Hershel is insufficient to resolve the disk. This can be overcome through high spectral resolution, where one uses velocity resolved spectral profiles of lines to place the source of line emission in the disk using the fall-off in rotation velocity with radial distance from the star as predicted by Kepler's 3$^{rd}$ law. The Herschel HIFI spectrometer, which had sufficient spectral resolving power, did not have sensitivity sufficient to perform these velocity-radial distance inversions for more than a few stars. Even then it only operated at frequencies below 1.9 THz, or 157 $\mu$m wavelength while the most important tracers of the inner disk lie at significantly higher frequencies. The ALMA interferometer has both the sensitivity and spatial resolution to begin to resolve protoplanetary disks, but it only operates at frequencies lower than about 900 GHz where the preponderance of spectral lines probe low excitation gas. These low-excitation line profiles  will be dominated by the outer, cooler and less dense regions of the disk and the parent molecular cloud. 

The best spectral lines to probe the inner regions of the protoplanetary disk will be probes of high density (n $\geq 10^{4}$ cm$^{-3}$) and intermediate temperature (T $\sim$ 100 to 500 K) gas. Such spectral lines are typically found in the mid- to far-infrared, and include the 440 $\rightarrow$ 313 28.914 $\mu$m, 643 $\rightarrow$ 616 32.313 $\mu$m, and 651 $\rightarrow$ 624 34.987 $\mu$m rotational transitions of H\textsubscript{2}O, the 1$\rightarrow$0 112.073 $\mu$m, 2$\rightarrow$1 56.230 $\mu$m, and 3$\rightarrow$2 37.702 $\mu$m rotational transitions of HD, and the $^{3}P_{0} \rightarrow^{3}$P$_{1}$ 63.184 $\mu$m fine-structure line of [OI]. HIRMES was designed to specifically address these transitions with velocity resolution up to 3 km s$^{-1}$ (spectral resolving power $R \approx 10^{5}$). Since orbital velocities in the disk will be of the order 3 to 30 km s$^{-1}$, this is sufficient to spectrally resolve these lines and make the link to emission from specific regions in the Keplerian disk. HIRMES also employs direct detection bolometers with sufficient sensitivity to be background limited at these high resolving powers so that a wide variety of sources can be investigated.  
Furthermore, HIRMES also contains a grating spectrometer with resolving power $\sim$ 900 that enables detection of the 43-47 $\mu$m solid-state water ice features. This combination of instrument parameters enables unique investigations of protoplanetary disks including:

\begin{enumerate}
\item \textbf{Tracing the total gas content in protoplanetary disks.}  
HIRMES can trace the total molecular hydrogen gas mass (H\textsubscript{2}) in the disk via velocity resolved studies of the HD isotopologue in protoplanetary disks. [The pure rotational lines of H\textsubscript{2} are very difficult to observe since they trace warmer gas than is in the disk, and have very weak line transition strengths.] Much of the HD emission will arise in the outer regions of the disk where gas is still accreting. The gas mass derived from HD reveals the efficiency of transferring gas into planets, especially the gas giants. HD is not observable from the ground, and HIRMES brings a unique combination of resolving power and sensitivity not available with space-based platforms.
\item \textbf{Tracing the water abundance.}
HIRMES can provide uniquely sensitive observations of water in the disk in regions near the snow line and combined with our HD observations we can derive its relative abundance. The water abundance is important, as it's extreme depletion relative to rock on Earth compared with its astrophysical abundance has led to the theory that water was delivered to a dry Earth long after its formation \cite{albarede} \cite{chyba}. However, it is not known if this process is common for exoplanets in the habitable zones around their stars. The water lines that best probe the habitable zones lie in the 28 to 35 $\mu$m regime with upper-level energies of 700 to 1000 K so that they trace gas at temperatures as low as T $\sim$ 200 to 300 K. These lines are unique to HIRMES and trace water vapor down to the snow line.

\item \textbf{Oxygen abundances, photochemistry, and velocities in the disk} HIRMES will measure and spectrally resolve the [OI] 63 $\mu$m fine-structure line and several OH rotational lines probing the photochemistry and thermal balance in outer layers of the disk \cite{weiden}, as well as the abundance of these two important gas phase reservoirs for oxygen. In addition, the [OI] line is particularly strong | it is expected to be four times brighter than the water or HD lines | so that substantial improvements in orbital dynamics will be made though its measurement, which as for the water and HD lines can be inverted through Kepler's laws to reveal spatial structure that is not available from direct imaging. 

\item \textbf{Water ice abundance and processing}. HIRMES's grating spectrometer mode delivers a complete, low resolution ($R \approx 600$) spectrum for protostars from about 25 to 120 $\mu$m. Of particular interest are the water ice feature near 45 $\mu$m, where HIRMES can spectrally resolve the solid state profiles of crystalline ice (peaking near 43 $\mu$m) and amorphous ice (peaking near 47 $\mu$m). The water ice content of the disk is derived from these features and the relative crystalline/amorphous abundance ratio provides information about its thermal evolution. In addition, the HIRMES grating spectrum can reveal the ice/rock ratio in protostellar disks \cite{McClure2015}. Is ice the dominant solid mass reservoir as in solar system comets, or is rock more important in the core-accretion phase of giant planet formation? 

\end{enumerate}

HIRMES will be capable of observing hundreds of protoplanetary disks from molecular clouds within 500 parsecs (pc) and over 100 in the three nearest associations at a distance of 140-160 pc from the solar system. With these spectral probes, HIRMES investigates the evolution of protoplanetary disks. We are particularly interested in the abundance and spatial distribution of water, and water ice at the time of planetesimal formation since it is these parameters that determine how much water can be delivered to worlds that can potentially support life.

\subsection{Fine-structure line studies}

As for the Hi-res, and Grating modes of HIRMES, the Mid-res and Low-res modes will have applications for a wide variety of science cases. In the Mid-res mode, a long (16 beam) slit is employed, and the Low-res modes use an imaging array of 16 $\times$ 16 pixels which greatly improve mapping speed. Here we focus on our primary drivers for these modes, which is mapping of the far-infrared line emission from both Galactic star formation regions, and over large scales in nearby galaxies. Our primary lines in these studies are the [OIII] 52 and 88 $\mu$m, the [NIII] 57 $\mu$m and [NII] 122 $\mu$m line which arise from ionized hydrogen (HII) regions formed by the ionizing radiation from nearby by O/B stars, and the [OI] 63 $\mu$m line which arises from the warm, dense photodissociation regions (PDRs) that form on the surfaces of molecular clouds exposed to far-UV (6 eV $\leq$ h$\nu$ $\leq$ 13.6 eV). These lines are quite bright and important coolants of the gas and therefore probe the sources of gas heating, which is in most cases stars. The lines and their ratios probe both the physical conditions of the gas \cite{staceyIEEE}.  The [OIII] 52 $\mu$m/88 $\mu$m line ratio is sensitive to HII region gas density, the [NIII] 57 $\mu$m/[NII] 122 $\mu$m and the [OIII] 88 $\mu$m/[NII] 122 $\mu$m ratio are sensitive the hardness of the stellar radiation fields that formed the HII region, the [NIII] 57 $\mu$m/([OIII] 88 $\mu$m + [OIII] 52 $\mu$m) line ratio is sensitive to the gas phase N/O abundance. The [OI] 63 $\mu$m line is sensitive to PDR gas density, and together with the [CII] 158 $\mu$m line yields PDR gas density and the strength of the ambient far-UV radiation field. Together these lines constrain many of the important physical parameters associated with star formation in both the Galaxy and at larger scales in external galaxies. Key questions include: 
\begin{enumerate}
\item $\textbf{What is the age of the current stellar population?}$ The age of the stellar population reveals the time since the last major star formation event. Star formation is likely stimulated by mechanisms that compress the parent molecular clouds such as gas orbit crowding at molecular bar/spiral arm interfaces, passage through spiral density waves, compression of gas by nearby supernova events, or galaxy-galaxy collisions. Therefore, to find the age of the stellar population is to probe the importance of such events to star formation. On galactic scales, the most massive star on the main sequence typically dominates the local radiation field. The most massive star will have the hottest photosphere (hardest radiation field) and shortest main-sequence lifetime, so that the [NIII]/[NII] and [OIII]/[NII] line ratios, which trace the hardness of the stellar radiation fields yield the time since the last star formation event. The former ratio has a residual gas density effect, typically removed with the [OIII] 52/88 line ratio, while the later ratio has a residual O/N abundance effect (see item 3).  

\item $\textbf{What is the intensity of current day star formation?}$ [OI] line imaging, together with [CII] line imaging (from Herschel/PACS or FIFI-LS on SOFIA) constrains both gas density and the strength of the ambient FUV radiation fields. The strength of the field is related to the numbers of stars per unit volume near the molecular clouds so that it is a measure of the intensity of the star formation event and the star formation efficiency, i.e. the fraction of a cloud mass that ends up in each generation of stars.

\item $\textbf{How many star formation events have occured?}$ Oxygen is a primary element, formed by the fusion of carbon and helium in the cores of massive stars.  Oxygen is released from the cores in the explosion of massive stars in supernovae events.  Nitrogen is a secondary element, formed predominantly in the CNO cycle that fuses hydrogen into helium in stars more than about 1.3 times the mass of the sun. Nitrogen is released to the interstellar medium through dredge-up and gas outflow events that occur as these intermediate stars enter their post main sequence red giant phase of evolution.  Therefore, nitrogen is enhanced every time a stellar population ages to the point that these type of stars become red giants. This time-frame is about 3-5 billion years.  Each generation of stars will therefore add more N relative to O, so the N/O ratio is a measure of the numbers of generations of stars that have occurred within a star formation region, or section of a galaxy. This then relates to star formation triggers and efficiency, thereby constraining star formation models. 
\end{enumerate}

\begin{figure} [ht]
   \begin{center}
   \includegraphics[height=6cm]{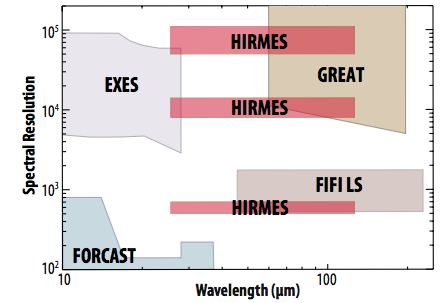}
   \end{center}
    \caption[disk] 
   { \label{suite} HIRMES provides SOFIA with access to new scientifically important wavelengths with high resolving power and sensitivity. For our science, HIRMES's sensitivity outperforms GREAT and FIFI-LS in their overlap region.}	
   \end{figure}

In addition, the HIRMES Mid-Res mode provides access to a variety of fine structure lines covering the wide range of ionizations states generated by fast interstellar shocks that may occur in fast molecular outflows, supernova events, or galaxy-galaxy collisions.  These lines include the fine-structure lines of [SI] 25.25 $\mu$m, [FeII] 25.99 $\mu$m, [SIII] 33.48 $\mu$m, [SiII] 34.81 $\mu$m, and [NeIII] 36.0 $\mu$m and rotational transitions of OH and CO. Line ratios probe shocked gas abundances, ionization state, and density in the shocked gas yielding mass flow estimates to be tested against shock models, while the line profiles that will elucidate the kinematics of the shocked gas \cite{stacey}.  

\subsection{HIRMES niche on SOFIA}

HIRMES enables SOFIA to open up new scientifically valuable regions of resolving power-wavelength phase space (Figure \ref{suite}). HIRMES provides new resolving power capabilities (R = 10$^{5}$, 1.2 $\times$ 10$^{4}$, 2000, and 600) at wavelengths from 28 to 63 $\mu$m, and continuous spectroscopy from 25 to 120 $\mu$m. Due to the use of bolometers and a 16 $\times$ 16 spatial array, HIRMES is significantly faster than FIFI-LS for mapping programs in the [OIII], [OI], [NIII], or [NII] lines. Due to the use of background limited bolometers, HIRMES is also more sensitive than the GREAT heterodyne receiver for high resolution spectroscopy of point sources.

\section{Fabry-Perot Interferometer Design: Optical}

\subsection{FPI Basics}
An FPI essentially consists of two highly reflective plane-parallel mirrors that form a resonate cavity.  Wavelengths that meet the resonance conditions will get transmitted due to constructive interference, while other wavelengths are eliminated due to destructive interference. The assembly of the two plane-parallel mirrors that form an FPI is also often called ``etalon''.

The resonating condition of an FPI is given by the equation: $2 \cdot d \cdot \cos (\theta) = m \cdot \lambda$, where $d$ is the separation of the mirrors, $\theta$ is the angle of incident, $\lambda$ is the wavelength, and $m$ is the order (integer) of the FPI. Therefore, for beams that pass the FPI at an angle the peak wavelengths are shifted to the blue with respect to the beam passing the FPI along the normal axis.

Given this resonant condition, it is clear that an FPI with a specific mirror separation can transmit a comb of wavelengths: the separation between the transmission peaks is called the free spectral range (FSR). In wavelength space (i.e. fixed mirror separation), the FSR is given by: $FSR = \frac{\lambda ^{2}}{2 \cdot d}$, while in ``mechanical space'', i.e. a given wavelength at varying mirror separation, the FSR is given by: $FSR = \lambda / 2$.
The figure of merit for an FPI is the finesse ${\cal F}$, which is the ratio of the free spectral range (FSR) and the full width half maximum (FWHM) of a transmission profile. Equivalently, the finesse can be viewed as the approximate number of ``bounces'' of the rays between the mirrors. In an ideal case the finesse is determined just by the reflectivity, ${\cal R}$, of a mirror: ${\cal F} = \frac{\pi \cdot \sqrt({\cal R})}{1 - {\cal R}}$. The spectral resolution of an FPI, $R$, is given by: $R = m \cdot {\cal F}$. Since the FPI is basically a resonator, the transmission profile has a Lorentzian shape.

\subsection{FPIs for HIRMES}

\begin{figure}%
    \centering
    \subfloat[Transmission vs. Finesse]{{\includegraphics[width=7.4
    cm]{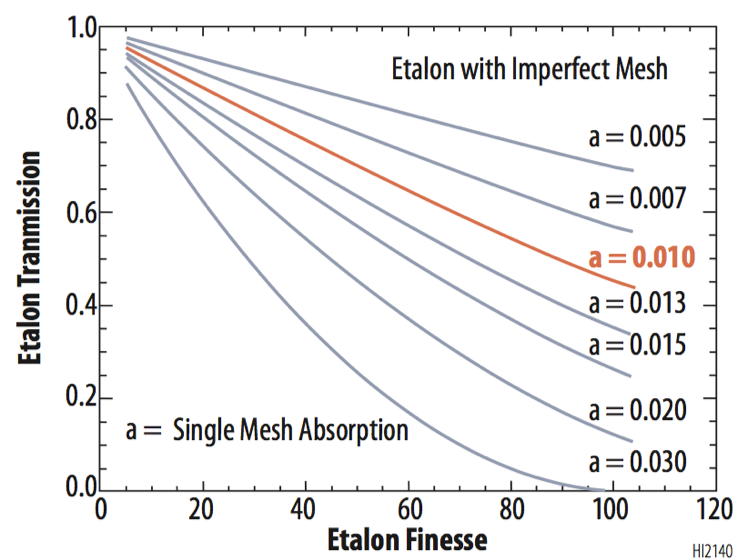} }}%
    \qquad
    \subfloat[Achievement of R=101,000]{{\includegraphics[width=7.4cm]{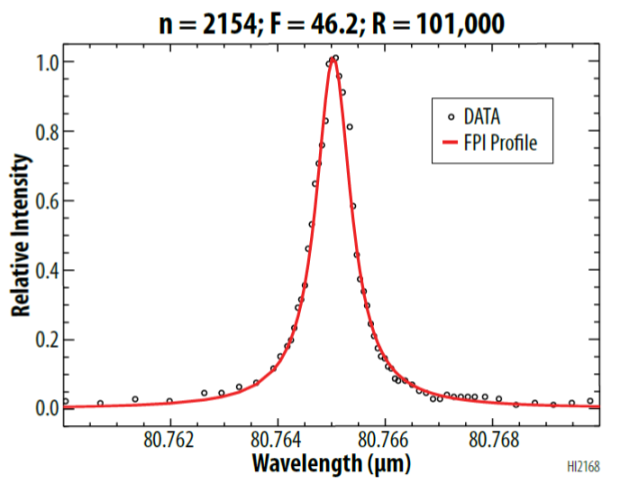} }}%
    \qquad
    \caption{ (left) FPI Transmission versus Finesse with various absorption coefficients.(right) Recent demonstration of $R=10^5$ using a 80.765 $\mu$m quantum cascade laser line.}%
    \label{abs}%
\end{figure}

For FPIs used in the far-infrared and longer wavelengths, free-standing metal meshes are usually used as the mirrors. A disadvantage of using simple metal meshes in an FPI is that the finesse increases with wavelength roughly as $\lambda^2$. While a high finesse benefits separating transmission peaks the etalon transmission decreases with increasing finesse for real meshes due to ohmic losses. Figure \ref{abs} (left) plots the FPI transmission versus finesse for various mesh absorption coefficients.  The optimal finesse for high-resolution FPIs is about 50 which yields a transmission coefficient of $\sim 0.7$.  Also plotted in Figure  \ref{abs} (right) is our first lab demonstation of resolving power of $10^{5}$ obtained with a low sensitivity detector and a 81 $\mu$m quantum cascade laser. 

\begin{figure}%
    \centering
    \subfloat[High Resolution Filter Wheel]{{\includegraphics[width=6.3cm]{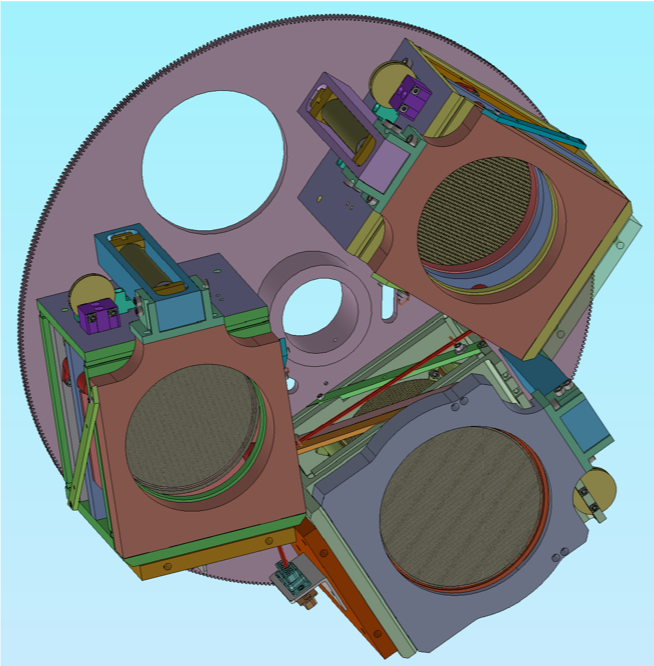} }}%
    \qquad
    \subfloat[Mid Resolution Filter Wheel]{{\includegraphics[width=7cm]{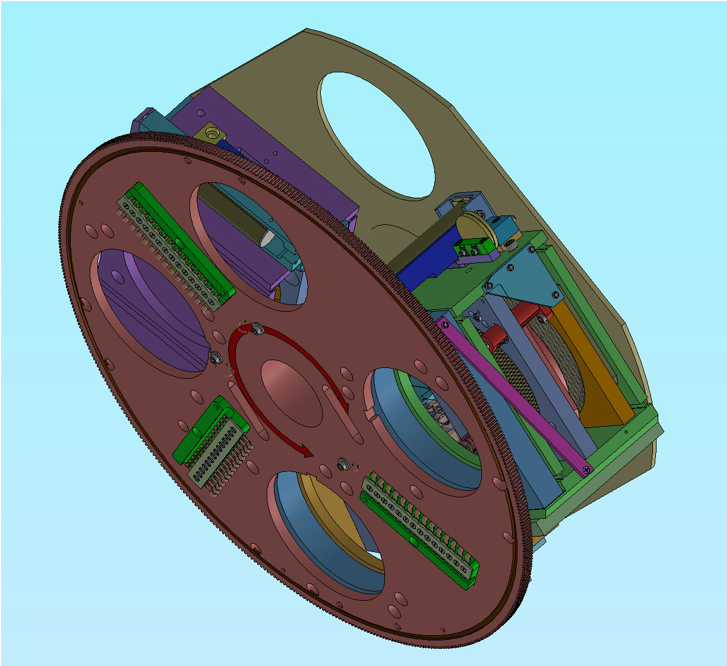} }}%
    \caption{ The filter wheel rotates the long, mid, and short wavelength FPIs or no FPI into the optical path allowing for simple and flexible mode selection. A rotary switch is used to simplify the electrical connection.}%
    \label{wheels}%
\end{figure}

Since HIRMES covers a wavelength range between 25~$\mu$m and 122~$\mu$m we employ three FPIs with different meshes, each FPI covering a subset of the total wavelength range. The meshes used in the FPI are from Precision Eforming and we use of the shelf nickel meshes with gold flash with 1000~LPI (lines per inch) for long wavelengths and 1500 LPI nickel for short wavelengths, as well as custom made 1200~LPI meshes for mid-wavelengths. To select a specific FPI for a given wavelength, the FPIs are mounted on a wheel, with three positions for the FPIs and an open position. Figure \ref{wheels} shows the Hi-res and Mid-res filter wheels. The rotary switch (discussed below) can be seen in green in the right hand model.

For its high spectral resolution spectroscopy (Hi-res) mode with $R = 10^{5}$ HIRMES uses FPIs. At such a high resolution, the FSR of an FPI is small and care must be taken remove unwanted spectral transmission profiles. To eliminate these unwanted orders we use FPIs with resolutions of 12,000 (the Mid-res FPIs) and a first order grating together with additional bandpass and blocking filters along the optical path. Thus the set of FPIs for Hi-res observations with HIRMES consist of three Hi-res FPIs ($R=10^{5}$ and $R=5 \times 10^{4}$ for the short wavelength) and three Mid-res FPIs ($R=12,000$).

\begin{figure} [ht]
   \begin{center}
   \includegraphics[height=5cm]{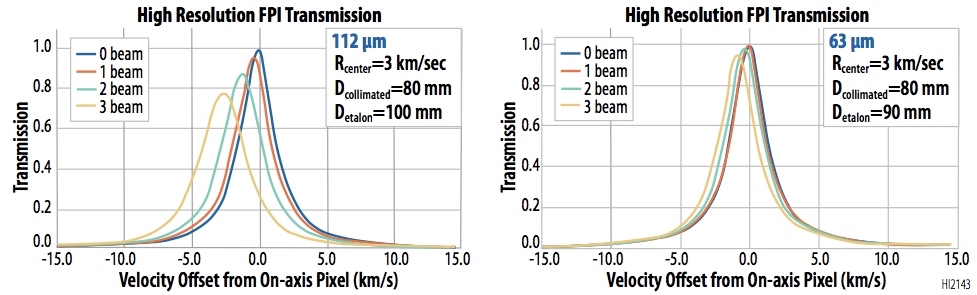}
   \end{center}
    \caption[disk] 
   { \label{hitrans} Hi-res FPI transmission profile at 112 $\mu$m (left) and 63 $\mu$m (right) for a $R = 10^5$ (3 km/s) for beams on the optical axis and 1, 2, and 3 beams off the optical axis.}
   \end{figure}

The HIRMES design calls for small spectral shift for off-axis detector pixels, with a spectral shift of less than a resolution element within the inner three detector pixels. This requirement leads to placing the Hi-res and Mid-res FPIs in a collimated beam near the pupil and a beam diameter at the pupil of about 8~cm. The clear aperture of the Hi- and Mid-res FPIs was then designed to be 9~cm.
However, to minimize the effect of ``walk-off'' (i.e. off-axis rays traveling off the mirror at each reflection) the clear aperture of the Hi-res long-wavelength FPI, which has the largest mirror separation, was chosen to be 10~cm. Figure \ref{hitrans}, demonstrates the modest reduction in both resolving power and transmission we expect at 112 and 63 $\mu$m for the Hi-res spectral profile even at 3 spatial postions off the optical axis of the system.

In addition to the high-resolution spectroscopy mode, HIRMES also includes a lower resolution imaging spectroscopy mode. For these observing modes, HIRMES uses low-resolution (Low-res) FPIs ($R=2000$) and first-order FPIs ($R\approx40$) to properly sort the orders of the Low-res FPI.  Due to the modest spectral resolution, the Low-res imaging FPIs need only have a clear aperture of 32~mm and can be mounted near a 20-m diameter pupil position. This pupil position acts as a first cold-stop inside HIRMES. The Low-res FPIs are thus ``miniature'' designs of the mid- and high-resolution FPIs. HIRMES employs two Low-res imaging FPIs that are mounted on a wheel. A separate wheel close to the 20~mm cold stop houses five first-order FPIs.

Except for the first-order FPIs, which have a fixed mirror separation, all FPIs in HIRMES can be scanned over a minimum of 1.5 FSR.

\begin{figure} [ht]
   \begin{center}
   \includegraphics[height=4cm]{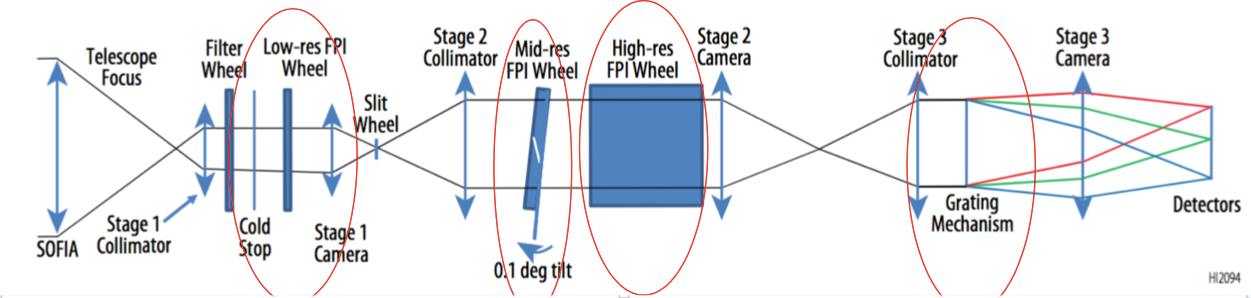}
   \end{center}
    \caption[disk] 
   { \label{optics} Block diagram of HIRMES optical path: the beam passes through a thermal blocking filter to the Stage 1 collimator; It then travels through the low-res FPI wheel, a four position slit wheel, and through the Stage 2 collimator; Then it passes through mid-res FPI wheel into the high-res FPI wheel. It continues through the Stage 3 collimator into the grating wheel, and then onto the two detector arrays. Baffles and separated stages serve to mitigate stray light. }
   \end{figure}

Figure \ref{optics} is a block diagram of the optical path required to address the various observing modes with HIRMES.  Within this diagram, one can see the tilt we have introduced between the Hi-res and Mid-res FPI  to eliminate parasitic FPI resonances due to rays bouncing between the etalons. 

\section{FPI Design: Mechanical}

The mechanical design of the low-, mid-, and high-resolution FPIs is based on a flex-vane parallelogram construction. Such designs were successfully used in several generations of FPIs used in ground-based and air-borne instruments \cite{poglitsch, latvak, bradford}. For the HIRMES FPI we have modified our previous designs in that the entire flex-vane stage was cut out of a single block of aluminum via wire-EDM (Electrical Discharge Machining). This fabrication method eliminates the difficult assembly of an FPI made of individual parts to the tight tolerances necessary for proper scanning. Detailed FEA (finite element analyses) were made for each of the FPIs to minimize any deformation of the flex-vane when scanning the FPI.

The free-standing metal meshes used as the FPI mirrors are stretched onto rings made out of magnetic stainless steel and are magnetically held in place on three point mounts in the flex-vane.
The first order FPIs deviate from this design, as these rings are made of stainless steel and clamped directly onto a wheel.

\subsection{High- And Mid-Resolution FPIs}

\begin{figure}%
    \centering
    \subfloat[Hi-res Long Wavelength FPI]{{\includegraphics[width=8.2
    cm]{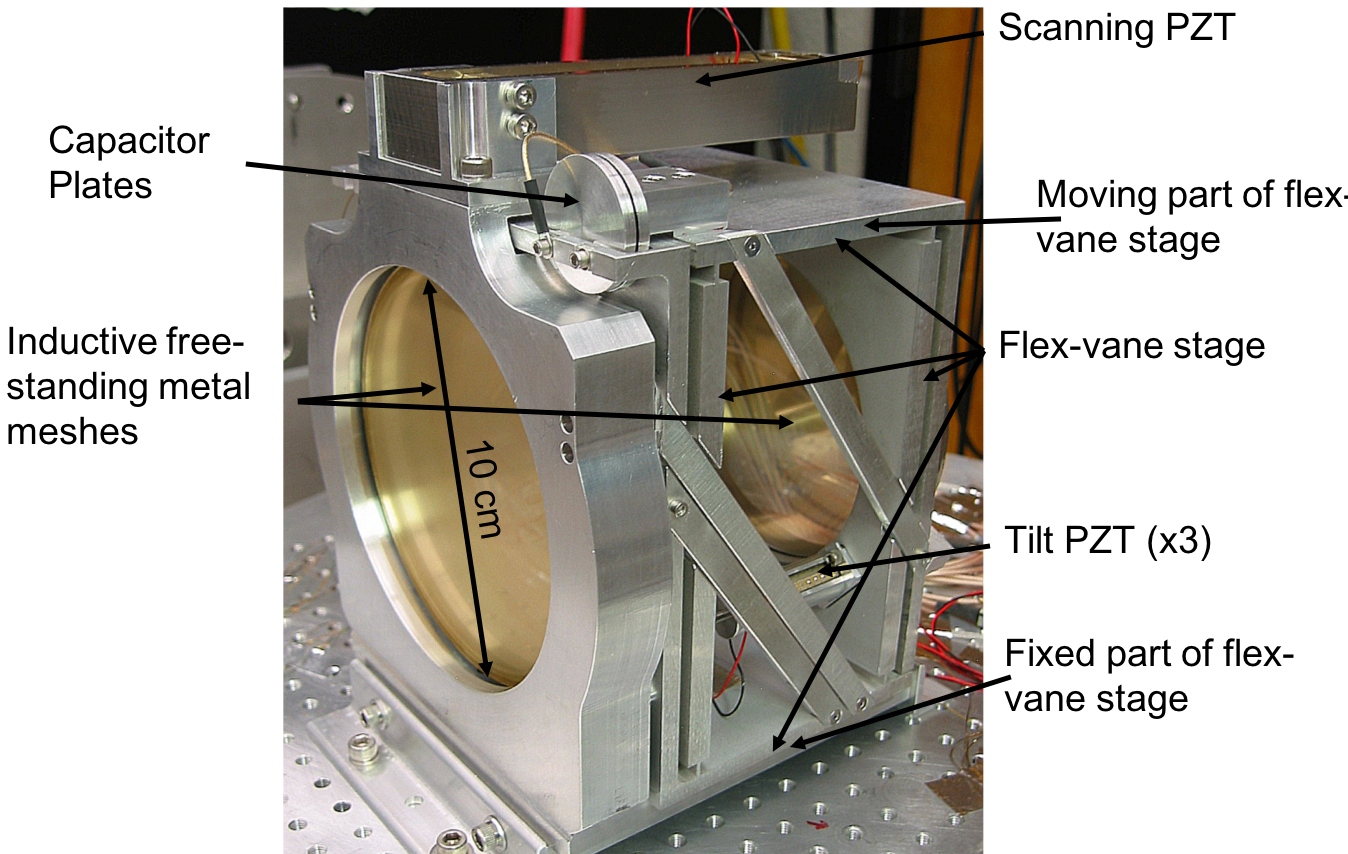} }}%
    \qquad
    \subfloat[Mid-res FPI]{{\includegraphics[width=3.9cm]{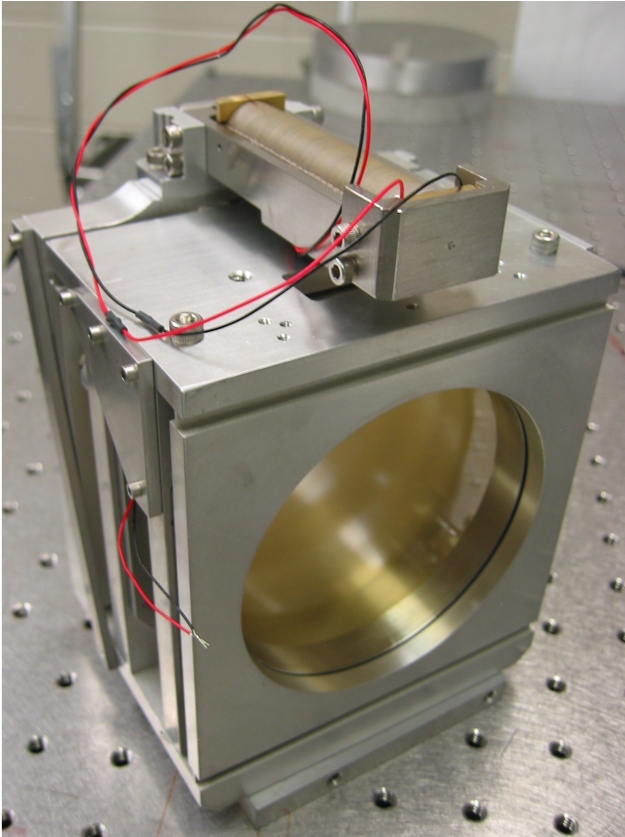} }}%
    \qquad
    \caption{ (a) Fully assembled Hi-res FPI with two free standing metal mesh mirrors on a flex-vane stage. 3 tilt PZTs and a scanning PZT control the mirror spacing and parallelism. (b) Partially assembled Mid-res FPI with scanning PZT shown.}%
    \label{high}%
\end{figure}

At present (July 2018) we have made and tested all three Hi-res FPIs at 77~K and two Mid-Res FPI at room temperature, and all of them showed no loss of mirror parallelism when scanned over more than 100 - 250 $\mu$m. Figure \ref{high} shows images of two of the HIRMES's flight Hi-Res (left) and Mid-res (right) FPIs.

To scan the Hi- and Mid-res FPIs and initially parallelize the mirrors we use low-voltage (100~V)  PZTs from Physik Instrumente. For scanning we use the P-088.781 PICMA PZTs and for parallelizing/tilting we use P-080.391 PICMA PZTs, which have travel range of 70~$\mu$m and 25~$\mu$m at room temperature, respectively.
Each of the Hi- and Mid-res FPIs uses four PZTs. Physik Instruments customized a $19''$ crate (E-500K136) to house four custom PZT amplifier modules (E-508K028). The custom PZT modules are high-voltage PZT amplifier that were modified to operate bipolar in the voltage range between -120~V to +120~V.

At cryogenic temperatures, these PZTs are predicted to travel about 24\% of their room-temperature travel when operated bipolar (-100~V to +100~V). We have mounted a P-088.781 PZT on our engineering version of a HIRMES FPI and cooled the FPI to 4~K. The travel of the PZT exceeded the prediction. However, it was questionable whether the PZT itself actually reached 4~K, and it is more likely that the PZT was at a temperature between 20 and 30~K.

Since the PZTs lose travel range when cooled to cryogenic temperatures, we employ a mechanical motion multiplier for the scanning PZTs to increase the travel. The motion multiplier is based on a simple lever-arm design using ball-bearings at the fulcrum and at the contacts to the PZT and the push-position at the flex-vane stage. The motion multipliers are designed with different multiplication factors depending on the necessary wavelength coverage and achieve factors between 4 and 9. However, due to the motion multiplication it is important to compensate for the different thermal expansions of the PZT and the flex-vane stage. Our design uses a mount made of Invar in which the PZTs are bracketed by blocks of brass. The compensation for thermal expansion is also crucial because we use capacitor plates to measure the relative displacement of the FPI mirrors and we need to ensure the capacitor plates do not touch when cooling the FPI.

Fine-adjustment screws with a fine thread providing a displacement of 200~$\mu$m per revolution are used to mechanically pre-adjust the parallelism of the FPI mirrors.

At this time, we have fully tested all three high-resolution FPIs at 77~K and delivered them to the HIRMES team at Goddard Space Flight Center (GSFC).

\subsection{Low-Resolution FPIs}

\begin{figure}%
    \centering
    \subfloat[Low Res FPI CAD Design]{{\includegraphics[width=5.4
    cm]{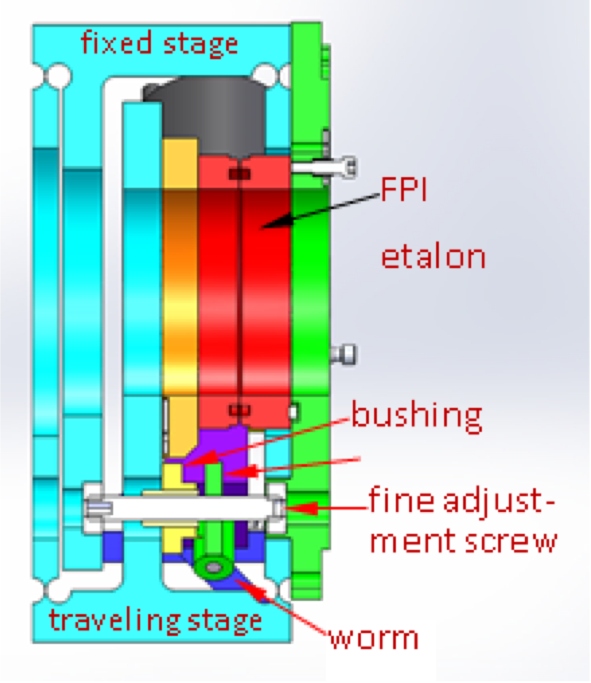} }}%
    \qquad
    \subfloat[Low-Res FPI flex vane stage]{{\includegraphics[width=4.2cm]{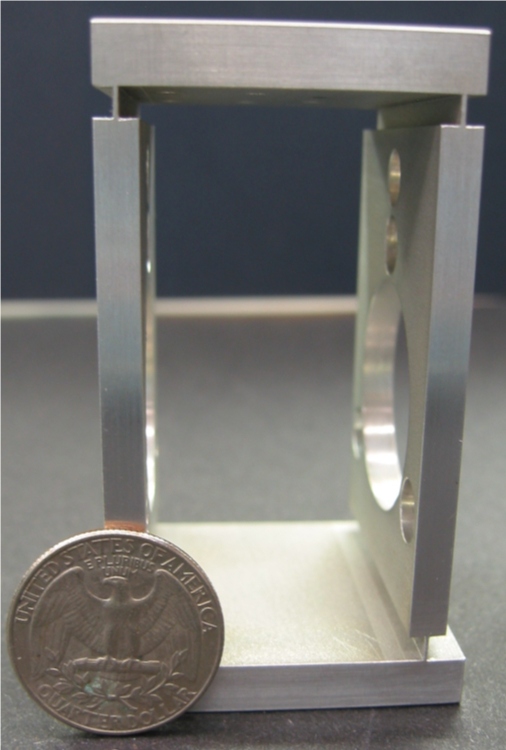} }}%
    \qquad
    \caption{ (a) The Low-res design has closer mirrors than the Mid-res and Hi-res and is controlled via a stepper motor and fine adjustment screw rather than by PZTs. (b) The Low-Res FPIs are still under assembly. Shown is a recently machined flex-vane stage.}%
    \label{low}%
\end{figure}

The two imaging Low-Res FPIs have a clear aperture of 32~mm. Their wire-EDM'ed flex-vane stage is basically a ``miniature'' design of the high- and mid-resolution FPIs. See Fig. \ref{low}.
However, instead of PZTs, these FPIs use cryogenic Phytron stepper motors (VSS19) to scan the flex-vane.
The stepper motors are required because these imaging FPIs need to scan over much larger range of up to 2~mm.
Due to the lower resolution and therefore the reduced constraints on finesse (i.e. parallelism) these FPIs do not use PZTs to adjust/compensate for tilt in the mirrors.
Instead, they use only fine adjustment screws to preset the parallelism of the mirrors.
Subsequent cool-downs of these FPIs will provide sufficient knowledge to properly preset the mirror parallelism.

\subsection{First-Order FPIs}

\begin{figure} [ht]
   \begin{center}
   \includegraphics[height=5.5cm]{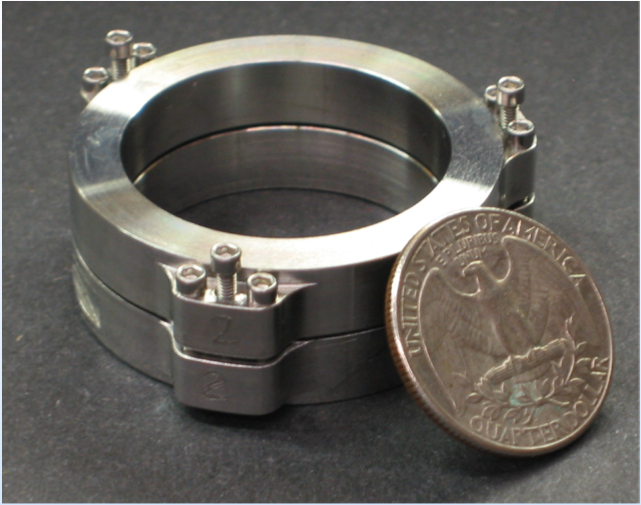}
   \end{center}
    \caption[disk] 
   { \label{fixed} Image of a meshless first-order FPI with push screws shown.}
   \end{figure}

The first-order FPIs consist of just two rings with meshes stretched onto them and the rings being separated by a ``push-pull'' screw design (See Fig. \ref{fixed}). We have used this type of design extensively, and it is based on the original design for the UCB Tandem FPI \cite {lugten}. The FPIs have a clear aperture of 32~mm and the rings are fabricated out of stainless steel.

\subsection{Rotary Switch}

Since each FPI is controlled by four PZT or a motor and includes a capacitor sensor, a thermometer and a limit-switch (for the low-resolution FPI), numerous wires would be required to operate all FPI and thus significantly increase the heat load into HIRMES.
We therefore employ a mechanical rotary switch the connects each FPI when in use to the external control electronics, and each wheel that contains FPIs is equipped with a rotary switch.
A rotary switch consist of several G10 blocks with contacts.
Each FPI has a contact block mounted the wheel and a fixed contact block mounted on the optical bench of HIRMES.
When each FPI on a wheel is rotated into the operation position, the corresponding rotating contact block connects to the fixed contact block on the optical bench.
This design therefore only requires external electronics and wiring to operate a single FPI, i.e. one external control electronics for all low-resolution FPIs, another control electronics for all mid-resolution FPIs and similarly for all high-resolution FPIs.

\section{Calibration}

The absolute position of the Hi-res and Mid-res FPIs can be determined through measurements of just two (non-resonant) far-IR lines in the lab. Once the absolute position is determined, it is calibrated against the capacitive sensor so absolute spacing during a scan can be maintained. In a power off situation, the FPIs relax to the zero position determined by the no-voltage applied PZT. Once absolute position is determined, regaining an adequate recalibration requires only a single spectral line, as the FPI can have moved no more than a faction of a FSR in the power-off scenario, at most, this would be by one resonant order from the power-on mode. The HIRMES cryostat also carries two far-IR Quantum Cascade lasers (QCL) (3.8 THz, 4.7 THz) to quickly enable FPI mirror precise spacing determination. The line widths of the lasers are narrower than 0.1 km/s and drifts are negligible. In addition, spectral calibration can be checked in the lab against gas cell absorption lines, and in-flight against telluric lines. The laser lines are bright enough to ensure a fixed fiducial wavelength is available to calibrate every HIRMES system mode.

Getting absolute calibration in the lab after an initial cool down can be completed quickly. The absolute spacing of both FPIs in a tandem FPI system can be found within a few hours of first getting the instrument cold enough to measure far-IR signals \cite{latvak}.
The FPI cavities can be tuned for parallelism of the mirrors by peaking the laser line intensity against the scan position and tilt PZT positions. Maximum signal, and minimum line width, occur for a properly-tuned cavity \cite{stacey}.

Prior to shipment to Goddard, the FPIs are tested mechanically at Cornell with expanded HeNe lasers at 78K. The criteria for success include parallelism during cool-down and during spectral scans, mirror flatness, scan length, and repeatability, stability, and reproducibility of metrology systems (capacitive bridge and stepper motor), and stability in the presence of changing gravity fields. All measurements are achievable solely with expanded HeNe metrology. Etalon efficiency and finesse is measured at room temperature within the FTS system at Cornell and at 4K within the FTS system at GSFC.

\section{Capacitance Sensor Bridge}

\begin{figure}%
    \centering
    \subfloat[Capacitance Bridge Circuitry]{{\includegraphics[width=5cm]{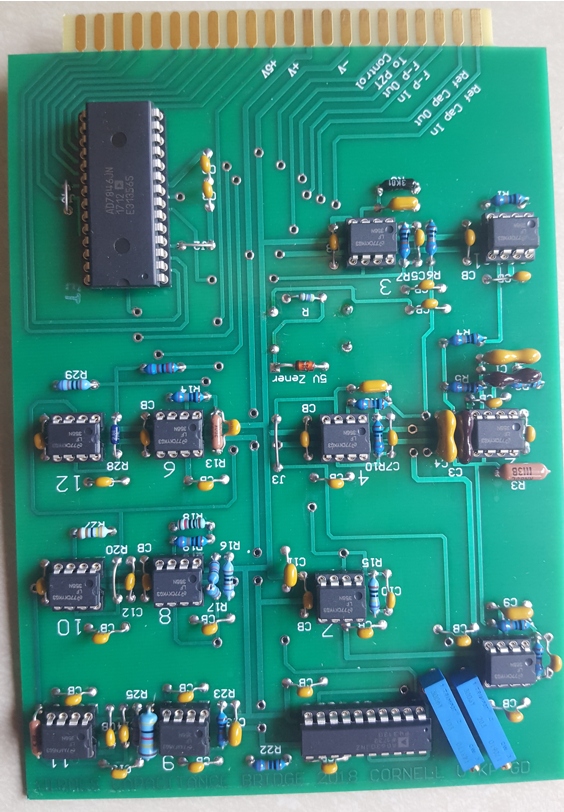} }}%
    \qquad
    \subfloat[Cap. Bridge Box (Outside)]{{\includegraphics[width=5cm]{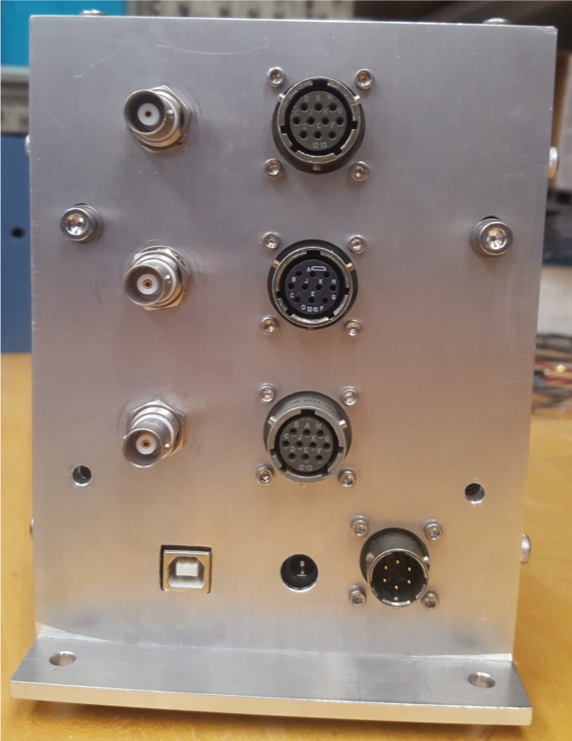} }}%
    \qquad
    \subfloat[Cap. Bridge Box (Inside)]{{\includegraphics[width=5cm]{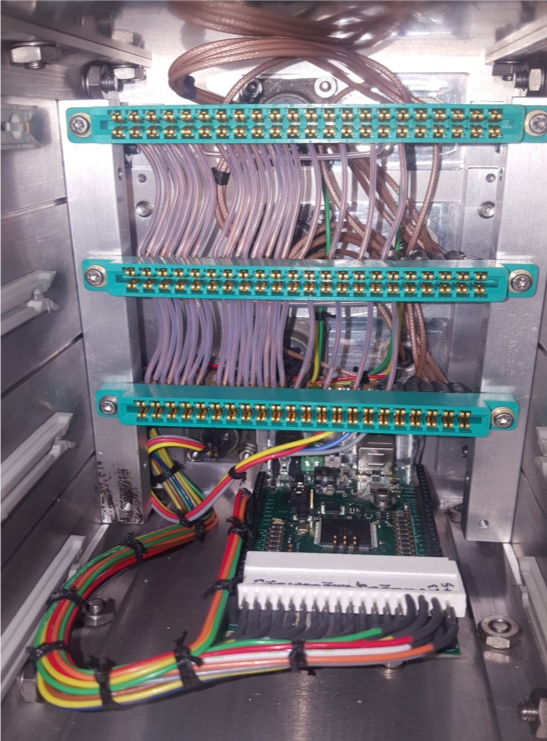} }}%
    \qquad
    \caption{ (a) PCB for sensing and controlling the scanning PZTs of the FPIs. (b) Front face of the capacitance bridge control box with FPI signal inputs, PZT controlling outputs, and an Arduino USB port. (c) Internal view of the control box showing edge connectors for the three bridges and the Arduino.}%
    \label{capbridge}%
\end{figure}

Since the PZT displacement is not a linear function of the applied voltage and actually shows a hysteresis for positive/negative voltage change, it is necessary to use a separate measure to track the FPI mirror separation. Even when employing a stepper motor to operate the FPI is wise to add a separate displacement measure since steps could be lost during operation, especially at cryogenic temperatures.

To measure the relative mirror displacement of the Lo-, Mid-, and Hi-res FPIs we employ a capacitance bridge circuit. On the flex-vane stages of each FPI is a capacitor consisting of two flat, parallel, 1.1 inch diameter disks. One of the disks remains fixed in position, while the other disk moves with the flex-vane and thus the scanning mirror. The gap between the capacitor plates is about 300 to 500~$\mu$m.
The capacitor bridge circuit compares the impedance of the capacitor on the FPI to the impedance of a fixed silver mica capacitor using a modest frequency AC-signal.
The signal to the FPI capacitor is 180$^{\circ}$ out of phase with the signal to the reference capacitor. If the bridge is perfectly balanced the combined signal passed through the capacitor would cancel out. A lock-in circuit compares the combined capacitor signal with the original oscillator signal to detect small imbalances and to generate an error signal. The error signal is smoothed and converted into a DC-signal that is feed back to the scanning PZT control input forcing the scanning PZT to adjust the mirror separation to balance the capacitor bridge.

To scan the FPI the sensor branch that sends the AC signal to the FPI capacitor includes a DAC (AD7846). This DAC allows to alter the amplitude of the sensor signal, thus driving the bridge out of balance, and subsequently forcing the scanning PZT to adjust the mirror and capacitor separation and hence balance the capacitor bridge.

This capacitor bridge design has been successfully been used in previous instruments \cite{poglitsch,latvak, bradford}. Figure \ref{capbridge} contains photographs of the capacitance bridge PCB and housing box. A difference from our previous instruments is that the reference capacitor in HIRMES will be mounted inside of the cryostat, as close as possible to the wheel to minimize effects due to the sensor cables. Inside the cryostat, the sensor cables are ultra-miniature, flexible coaxial cables. Stainless steel version of these coax cables are used when the cables cross different temperature stages of HIRMES. If the coax cables are only mounted on a single temperature stage the cable material is copper.

The AD7846 is a 16 bit DAC providing a resolution of 65536 of the control DC signal to the scanning PZT ($\sim -10$~V to $\sim +10$~V). The Physik Instrument E-508K028 PZT amplifier module amplifies a input control voltage by a factor of 12 to an output voltage to the PZT. For a finesse of 50 and $\lambda = 63~\mu$m the mechanical resolution element corresponds to $\Delta d = \lambda / (2 \cdot {\cal F}) = 630$~nm.
The FPIs should be able to scan in steps of 1/5 of a resolution element. To allow for proper wavelength setting of the steps the capacitor bridge needs to be able to set the mirror separation to within 1/10 of a resolution element. To continue with the example, assume that the FPI can scan 1.5 FSR at $63~\mu$m, or $\approx 48~\mu$m. For a bipolar PZT scanning voltage range of 200~V this then results in $48 / 200 = 0.24~\mu$m/V.
However, the control voltage from the bridge to the PZT is a factor 12 lower, resulting in $48 \cdot 12 / 200 = 2.88~\mu$m/V.  Thus a step size of 60~nm would require a bridge control voltage change of $0.06 / 2.88 \approx 0.02$~V. The cryogenic (77 K) scan tests we have conducted showed travel range per DAC bit of about 0.02~$\mu$m/bit.

Although the voltage resolution shown above appears easily achievable, noise and systematics could become an issue. To prevent cross-talk between the capacitor bridges that control the mid- and high-resolution FPIs and monitor the low-resolution FPI needs each of the bridge circuits uses a different oscillator frequency. For example, the oscillator frequencies for the mid- and high-resolution FPI capacitor bridges is 16~kHz and 50~kHz, respectively. We have measure noise in the bridge control voltage of order 5~mV using an oscilloscope, which most of the fluctuations appearing at high frequencies. A better noise performance analysis is ongoing. It is also imperative to provide proper grounding. The capacitor bridge PCB will be housed in an alodyned box mounted directly onto HIRMES, which will provide a good reference ground. In the lab, with both the capacitor bridge and the cryostat sharing a proper ground we have not seen systematic in bit/voltage offsets. These only occurred if grounding was removed.

The computer interface to the capacitor bridge circuits is via a Rugged Mega by the company Rugged-Circuits.
The Rugged Mega is based on an ATmega 2560, similar to an Arduino Mega 2560, and shares the same functionality and software sketches as an Arduino.
A single Rugged Mega controls all three capacitor bridge circuits for the HIRMES FPIs.

Our tests up to this time have shown that our bridge circuity has enough bit and voltage resolution to achieve a step size of 60~nm, and that it that noise appears a factor of about 4 below the lowest voltage step.
A capacitor bridge sensor PCB for the high-resolution FPIs with a Rugged Mega board in an alodyned crate was delivered to the HIRMES group at GSFC.

\section{Conclusion}

HIRMES's primary goal is to study key constituents of protoplanetary disks: water vapor and ice (which play a critical role in forming giant planet cores and producing habitable conditions on terrestrial planets), and neutral oxygen, a tracer of disk chemistry and radial structure \cite{kominami}. An important step is to measure hydrogen deuteride (HD), a tracer of molecular hydrogen, which will determine the protoplanetary disk mass with an accuracy not achievable by existing instruments. The high sensitivity and resolving power we will obtain with HIRMES will resolve narrow emission lines and determine their origin spatially in the disk from velocity information allowing for detailed disk modeling. In particular, the critical transition region demarcated by the snowline at a few to 10 AU will be solely accessible by HIRMES. This will allow HIRMES to help us learn more about the delivery of water to habitable worlds.

The Hi-res FPIs have already been delivered to NASA GSFC by Cornell (Aug. 2018). The Mid-res FPIs are scheduled to be delivered on Sept. 1, 2018, and the Lo-res FPIs are expected to be delivered by the end of September. First light is expected one year later, and HIRMES will become a SOFIA facility instrument in April 2020. Assembly of the complete optical bench is still on going at GSFC. The high-res FPIs have all been tested at Cornell at cryogenic temperatures (77 K) prior to delivery to GSFC using the (16 KHz) cap. bridge driven PZTs and have met all required specifications.
\section{Acknowledgements}
The HIRMES FPI work at Cornell University is supported by NASA grant 80NSSC18K104.
\bibliography{report} 
\bibliographystyle{spiebib} 

\end{document}